# On the Mechanism of Laminar-turbulent Transition and the Origin of the Critical Reynolds Number

Andrei Nechayev

*Moscow State University, Geographic Department*

*A theoretical mechanism of laminar-turbulent transition originated from the deceleration of fluid srteams on the walls of the channel or pipe is proposed. For Poiseuille flow an analytical expression relating the critical Reynolds number with the degree of disturbance of the flow is derived.*

Turbulence is among the most interesting and mysterious phenomena of classical physics. The main role of Turbulence performance a vortex plays. It represents the macroscopic (by comparison with molecular size) object formed by fluid being in specific rotational motion.

Transition to turbulence occurs in the laminar state in which layers of fluid move parallel to the axis of flow that is easily visualized using a thin colored stream which becomes entirely eroded by transverse movements when turbulence is fully developed. The fact of the laminar-turbulent transition and its critical condition were determined by Osborne Reynolds more than a hundred years ago [1,2] but adequate physical and mathematical description of this transition is still lacking [3]. Turbulence problem keeps the urgency despite the exponential growth in the quantity of publications devoted to this phenomenon [4]. In particular, we can't understand how the Navier-Stokes equation providing a full description of the fluid dynamics lead to the critical Reynolds number $\text{Re}_{cr}$ which makes a boundary between laminar and turbulent regimes.

Vortex is a "visit card" of turbulence. It is known that its formation is genetically linked with contrary flows. We assume that the appearance of local contrary streams in a laminar flow is a first sign of turbulence.

The existence of contrary flows is provided by oppositely directed pressure gradients. Why can they emerge in a laminar flow? Bernoulli's equation gives us a hint: the pressure gradient along the line current $\partial p / \partial s$ is proportional to the convective acceleration $v \partial v / \partial s$ with opposite sign. The pressure increases where fluid decelerating occurs. Braking takes place near solid walls if the fluid particles have the corresponding velocity component. Decelerating on the roughness of the walls, the streams can generate local zones of high pressure that are capable of deploying the stream (or its portion) in the reverse direction.

We demonstrate how the near wall decelerating of the deflected stream allows to obtain an analytic expression for the critical Reynolds number. Consider Poiseuille flow in a pipe (Fig. 1):

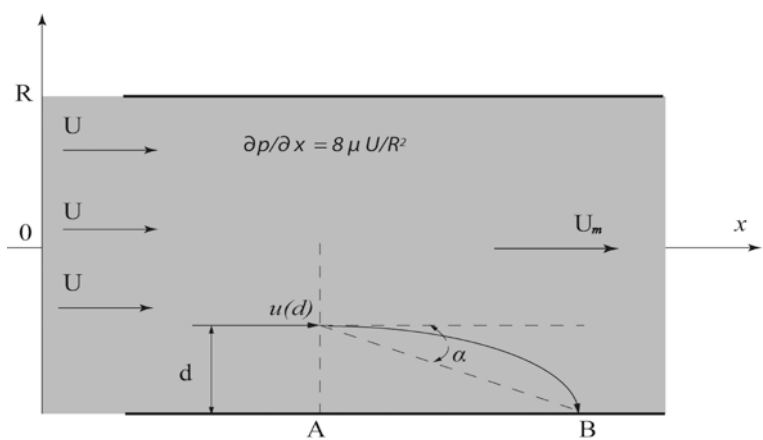

*Fig.1 Scheme of the laminar fluid flow in the pipe. Curved arrow denotes the stream deflected from the axis of the flow. **A** is a starting point of the deflection, **B** is a point of collision with the wall.*

Assume that some stream deflects the flow axis at the cross section $x = A$ at distance $d$ from the tube wall (Fig.1). We leave aside the problem of the genesis of similar streams (disturbance in the pipe input or wall vibration can serve as a cause of their formation), it is important that this stream has to touch the wall and its particles velocity in certain places becomes equal or very close to zero. In accordance with Bernoulli's law (viscosity is neglected) due to stream decelerating the pressure in the vicinity of point B can be increased by the value of $\rho[u(d)]^2/2$ relatively to the initial pressure which was less than the pressure at section A by an amount equal to $(\frac{\partial p}{\partial x}\frac{d}{tg\alpha})$. If the pressure increasing at the point B exceeds the flow pressure drop from point A to point B, the reverse pressure drop and reverse motion capable of the vortex formation is possible. Thus, the emergence of turbulence can be attributed to the following conditions:

$$\rho\frac{(u|_{y=d-R})^2}{2} > -\frac{\partial p}{\partial x}\frac{d}{tg\alpha} \qquad (1)$$

Using well-known formulas for the Poiseuille flow and considering $d = kR$ where $0 < k < 1$ we obtain:

$$\rho\frac{4U^2k^2(2-k)^2}{2} > \mu\frac{8U}{R^2}\frac{kR}{tg\alpha} \qquad (2)$$

where $U$ is the average velocity of flow, $\mu, \rho$ are the fluid viscosity and density.
In final form:

$$\frac{\rho R U}{\mu} \equiv \text{Re} > \frac{4}{k(2-k)^2 tg\alpha} \equiv \text{Re}_{cr} \qquad (3)$$

Thus, the condition (3) gives an analytic expression for the critical Reynolds number in the flow with one disturbance with given parameters ($k, \alpha$). According to it, the closer the deflected stream is to the wall (less its velocity), the higher is the critical Reynolds number. Similarly, the smaller is the deflection angle $\alpha$, the more is the critical number. The minimum value of $\text{Re}_{cr}$ is at $k = 1$ that is when the central stream is deflected. It is easy to see that for $k \ll 1$ (decelerating streams are very close to the wall) $\text{Re}_{cr}$ can grow indefinitely, regardless of the deflection angle. This angle, obviously, can't be more than 45° when transversal velocity is equal to $u$.

So each disturbed stream has its "personal" $\text{Re}_{cr}$ depending on the distance $d$ and the deflection angle $\alpha$. The $\text{Re}_{cr}$ of the total flow will be "selected" from all stream critical numbers and must be the minimum one. It is clear that for a stable turbulence the disturbances providing streams deflection must be consistently reproduced, otherwise (with the weakening of the disturbance) the turbulent and laminar flow will be intermittent. If the condition (3) is not satisfied, the pressure from A to B will decrease and the stream will be carried away downstream remaining regime to be laminar.

The above considerations allow us to construct a simplified picture of the laminar-turbulent transition. The result of laminar flow disturbance is the appearance of some streams with the transverse velocity component which determines the angle of deflection. But the main thing is the average velocity of streams that reach the walls and decelerate. If somewhere the criterion (1) is satisfied and the vortex formed, due to its volume it will disrupt the laminar flow pattern and deflect streams located closer to the flow axis with higher speed which enable them to overcome the condition (1) after decelerating near pipe wall. A sort of chain reaction begins: small primary vortices near the walls shifted downstream, drifting toward the center and provoking the formation of other, more "powerful" vortices. As a result, the flow proceeds to a new state, in which the viscosity plays a minor role, and the stability is ensured by the friction of macro vortices and jets against the walls of the pipe. Perhaps this strongly nonlinear flow state defines the minimum critical Reynolds number and gives the corresponding hysteresis on the curve $U(\partial p/\partial x)$ indicating that the reverse transition in the laminar state occurs at lower average velocity than the direct laminar-turbulent transition.

We note in conclusion that the above described mechanism of turbulence is not genetically linked to the viscosity of the fluid: in accordance with (1) it is determined only by the pressure gradient $\partial p/\partial x$, flow velocity profile and the presence of perturbations and therefore it can occur in non-Newtonian fluids and in superfluids.

## Conclusions

It is supposed that turbulence begins with the appearance of local reverse currents in the laminar flow. It is shown that such currents can emerge if some streams decelerate near solid wall of pipe or conduit. Each disturbed stream has its "personal' critical Reynolds number determined by two stream parameters: original distance from the wall and deflection angle. Exceeding this "individual" critical number results in the reverse flow near the point of stream-to-wall collision and possible formation of a vortex. The minimum of all critical numbers becomes critical Reynolds number for the flow as a whole.

## References

1. Reynolds O. On the dynamical theory of incompressible viscous fluids and the determination of the criterion. *Phil. Trans. Roy. Soc.* 186A, 123-164 (1895)
2. Schlichting H., Gersten K. Boundary Layer Theory. *Springer, Berlin, Heidelberg, New York,* 2000
3. Ecke R. The Turbulence Problem. *Los Alamos Science*. 2005, Number 29, p.124-141
4. Lumley J.L., Yaglom A.M. A Century of Turbulence. *Flow, Turbulence and Combustion*, 2001, 66, p.241-286